\documentclass[journal]{IEEEtran}
\usepackage{graphicx}
\usepackage{amsmath}
\usepackage{algorithmic}
\usepackage{array}
\usepackage[caption=false,font=footnotesize]{subfig}
\usepackage{url}
\usepackage{balance}

\ifCLASSINFOpdf
\else
\fi
\begin{document}
\title{Binocular Rivalry - Psychovisual Challenge in Stereoscopic Video Error Concealment}
\author{Md Mehedi Hasan,~\IEEEmembership{Member,~IEEE,}
        John F. Arnold,~\IEEEmembership{Senior Member,~IEEE,} and \\
        Michael R. Frater,~\IEEEmembership{Senior Member,~IEEE}
\thanks{M. M. Hasan is with PBRC, University of South Australia, SA 5095, AUS (e-mail:mdmehedi.hasan@unisa.edu.au).}
\thanks{J. F. Arnold and M. R. Frater are with SEIT, University of New South Wales, ACT 2610, AUS (e-mail: j.arnold@adfa.edu.au; m.frater@adfa.edu.au).}
\thanks{}}
\maketitle

\begin{abstract}
During Stereoscopic 3D (S3D) video transmission, one or both views can be affected by bit errors and packet losses caused by adverse channel conditions, delay or jitter. Typically, the Human Visual System (HVS) is incapable of aligning and fusing stereoscopic content if one view is affected by artefacts caused by compression, transmission and rendering with distorted patterns being perceived as alterations of the original which presents a shimmering effect known as binocular rivalry and is detrimental to a user's Quality of Experience (QoE). This study attempts to quantify the effects of binocular rivalry for stereoscopic videos. Existing approaches, in which one or more frames are lost in one or both views undergo error concealment, are implemented. Then, subjective testing is carried out on the error concealed 3D video sequences. The evaluations provided by these subjects were then combined and analysed using a standard Student t-test thus quantifying the impact of binocular rivalry and allowing the impact to be compared with that of monocular viewing. The main focus is implementing error-resilient video communication, avoiding the detrimental effects of binocular rivalry and improving the overall QoE of viewers.
\end{abstract}

\IEEEpeerreviewmaketitle

\section{Introduction}
\IEEEPARstart{3}{D} videos are becoming common in not only commercial cinemas but also home entertainment systems. The demand for high-definition 3D videos is growing rapidly in the mass consumer market and influenced by the investigation of 3D video Quality of Experience (QoE) \cite{wang2012perceived}, \cite{hewage2013quality}. Different artefacts can occur at each stage of the processing chain as shown in Fig.1, especially in the coding, transmission and display stages. Apart from the artefacts in different video processing steps, the transmission network itself often introduces errors due to delay or packet loss. The impacts of network errors and their concealment in 2D videos ensured that error-resilient video transmission was widely discussed in the past \cite{biswas2009improved}, \cite{zhang2000cell}.

However, the scenario is different for 3D videos as distortions in one or both views are perceived quite differently \cite{carreira2010subjective}, \cite{barkowsky2010subjective} and create detrimental effects which lead to binocular rivalry \cite{zhao2013rivalvy}, \cite{bensalma2013perceptual}. This strongly degrades the user QoE as it produces visual discomfort which can cause visual fatigue, eye strain, headaches and nausea \cite{lambooij2009visual}. Subjective analysis \cite{pinto2011subjective} has also shown that a similar packet loss rate can have a significantly different impact on a user's perception of quality for 3D compared to 2D video. Since the Human Visual System (HVS) is more sensitive to 3D perception, the impact of transmission errors after appropriate error concealment need to be carefully analysed.

\begin{figure}[!t]
\centering
\includegraphics[width=3.4in]{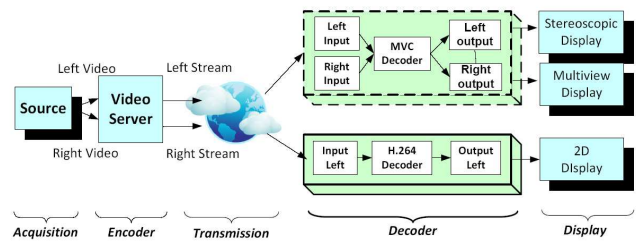}
\caption{Overview of 3D Video Processing Chain}
\label{fig_1}
\end{figure}

\begin{figure}[h!]
\centering
\includegraphics[width=3.5in]{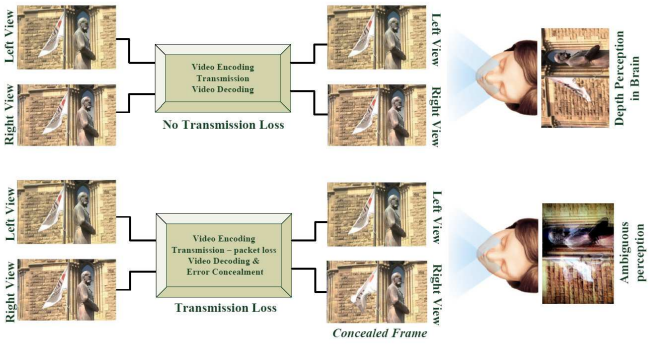}
\caption{Binocular Rivalry: perceptual ambiguity caused by transmission loss}
\label{fig_12}
\end{figure}

According to the suppression theory of binocular vision, there is only one form of binocular interaction, called rivalry \cite{zhang2011binocular}, which operates for both similar and dissimilar images. It is a visual phenomenon that occurs when dissimilar monocular stimuli are presented to corresponding retinal locations on the two eyes as shown in Fig. \ref{fig_12}. The binocular suppression theory \cite{lei1994spatial} suggests that, in these situations, the overall perception is usually driven by the quality of the better view (i.e., left or right) providing that the quality of the poorer view is above a threshold value. However, this capability is limited and studies have shown that additional cognitive load \cite{zhang2011binocular} is necessary to fuse these views and that this can lead to visual fatigue and eye strain, and prevent users from watching 3D content for long periods of time. This issue has resulted in many 3D TV channels, such as ESPN 3D, Foxtel 3D, N3D and 3NET ceasing to broadcast \cite{3dtv2016} as well as broadcasting services, such as MSG 3D and Sky 3D channels, being limited.

Other than the quantitative quality measures, very few studies have been conducted to subjectively analyze error concealed stereoscopic videos \cite{wang2012perceived}, \cite{carreira2010subjective}, \cite{barkowsky2010subjective}. Our recent research findings \cite{hasan2014subjective} showed that not only transmission losses but also applying conventional 2D error concealment techniques in a 3D video decoder triggers such artefacts and presents a perceptual challenge that requires extensive analysis to determine the psycho-visual impacts.

In this paper, we focus on these issues, i.e., we analyse an error concealment strategy for erroneous transmission of stereoscopic video with an appropriate subjective testing scenario. Based on the subject's Mean Opinion Score (MOS) on original video content, Least Square Regression Line (LSRL) analysis is performed to determine their non-deterministic behaviour and weaker correlation in scores while watching 3D video comparing to 2D that highlights the binocularity of the HVS. Through correlation between error concealment approaches, the sensitivity of this binocularity is analyzed and then the Student t-test \cite{rochowicz2012p} is used to compare 3D and 2D videos in order to define the statistical significance of binocular rivalry in the former when error concealment is applied independently to each view.


\section{Existing Error Concealment Approaches}

Most error concealment techniques found in the literature can handle macroblock (MB) or slice losses using neighbouring MB information, such as motion vectors (MVs) and pixel values \cite{micallef2010error}. However, in low bit-rate video coding, packet losses can result in the loss of a whole frame and may also lead to further frame losses before the decoder can re-synchronize to the bit-stream. In the case of frame loss, temporal and spatial interpolation or motion extrapolation within a frame do not work effectively. Also, due to limited time constrains in real-time applications frame-loss concealment is the most convenient way to conceal the lost frames.

\textbf{H.264 Concealment} \cite{micallef2010error} is a computationally intensive algorithm as it involves the sophisticated error concealment method implemented in the H.264 reference software and uses spatial or temporal interpolation depending on the frame type. However, in 3D video error concealment, because of distortions in either one or both views, binocular rivalry occurs as the error concealment artefacts are visible only in one view or are visible in both views but in different locations.

\textbf{Switching to 2D} \cite{wang2012perceived}, \cite{barkowsky2010subjective} is another approach in which a video is switched from a 3D to 2D representation when an error occurs in only one view. As the other view is undistorted, it is displayed to both eyes which leads to a 2D impression. As a result of their higher levels of disparity, 3D videos achieve more quality in terms of depth perception. However, switching from 3D to 2D and then from 2D to 3D creates high transitional distortions which diminish the overall video quality and the comfort level experienced by users, especially in high definition videos with significant depths which, in turn, lead to visual fatigue and discomfort. Therefore, there needs to be a trade-off between disparity and depth perception.

\begin{figure}[h!]
\centering
\includegraphics[width=3.5in]{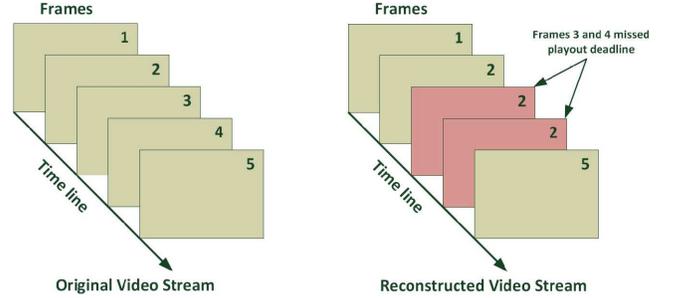}
\caption{Freeze Frame error concealment}
\label{fig_13}
\end{figure}

\textbf{Frame Freeze} \cite{wang2012perceived}, \cite{carreira2010subjective}, \cite{barkowsky2010subjective} is a widely used frame-loss concealment method for real-time video transmission and broadcasting and uses the last frame that was correctly received to replace all incorrect and missing frames. Fig. \ref{fig_13} shows a frame freeze error concealment approach in which the video decoder substitutes frame 2 for frames 3 and 4 when the latter are not received before their playout deadlines. Frame freezing mechanisms employed to tackle missing frames caused by transmission errors or delay in a single view can lead to temporal de-synchronization, whereby one eye sees delayed content compared with that seen by the other eye, thus creating binocular rivalry.

In \textbf{Double Freeze} \cite{carreira2010subjective}, \cite{pinto2011subjective} the last frame correctly received for both views is displayed after any transmission errors affecting either or both views. Frame freeze on single or double views produces good performance in low motion videos. For long duration videos, reduced playback speed is annoying and creates transitional distortions in high motion videos.

Finally, in \textbf{Reduced Playback Speed} \cite{barkowsky2010subjective} instead of completely stopping the video, it is assumed that there is a buffer of video frames that contains half a second (roughly 12 frames) of decoded content. As these frames are played back during the decoder's recovery time, a user observes that the playback slows down, skips and then continues at normal speed.

Errors in transmission over unreliable communication channels are greater for Stereoscopic 3D (S3D) than for 2D video as the former has two independent channels, each of which can suffer uncorrelated impairments. For instance, a delay in one view could cause temporal desynchronization, which can lead to a reduction in 3D viewing comfort. Also, the methods employed to mitigate these artefacts (e.g., error concealment) do not work as effectively for 3D as 2D videos  \cite{wang2012perceived}. Ideally, the two 3D channels need to be carefully synchronized to generate proper 3D depth perception and avoid binocular rivalry. Human vision research on the impact and modelling of the binocular rivalry of stereoscopic images or videos is still in an early stage, with the following two issues worth considering:
\begin{enumerate}
  \item how binocular rivalry affects the perception of 3D video quality and visual comfort; and
  \item which kinds of inter-view image differences can cause binocular rivalry.
\end{enumerate}

\begin{figure*}[!t]
\centering
\includegraphics[width=5.3in]{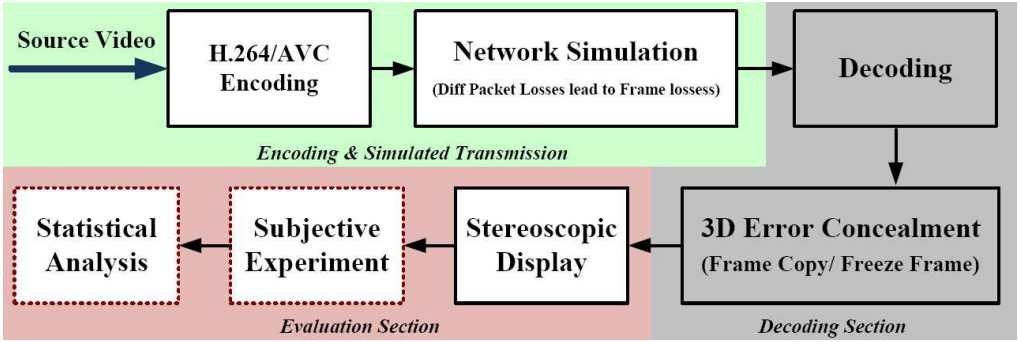}
\caption{Steps for evaluating transmitted and error-concealed 3D videos}
\label{fig_2}
\end{figure*}

\section{Proposed Method}
\subsection{Video Processing Chain \& Testing Conditions}

Subjective assessment is the most effective way of judging the perceived quality of a received video distorted by network impairments which are subsequently concealed. The video sequences for the subjective experiments were prepared in a simulated transmission chain, as shown in Fig. \ref{fig_2}. Several different scenarios, called Hypothetical Reference Circuits (HRC) according to the terminology of the VQEG \cite{cermak2009test}, were used to create these sequences which were encoded according to the standards used in the H.264 JM 18.6 \cite{marpe2006h} \cite{h264soft} reference software platform. The network simulation assumed that the loss of a single Round Transport Protocol (RTP) packet created a single frame loss which was then concealed by the widely used Frame Freeze error concealment technique at the decoder. The user experience was measured through both a subjective evaluation and analysis of the psychovisual impact.

For subjective testing, a 32 inch Samsung 3D TV was used to display each 3D video and an ACTIVE shutter glass made by Nvidia 3D vision used to watch it. According to ITU-R BT.500.13 \cite{itu500} and the VQEG HDTV test plan,the viewing distance was 3 times the height (3H) of the display which was positioned 1 meter from the wall to avoid conflict and flickering, with the room illumination adjusted to ensure that background light of no more than 15\% of the display's illumination was allowed. The wall behind the monitor was placed in such a way that the light did not fall on the viewers and did not exceed 5\% of the monitor's peak luminance (when functioning as a stereo monitor), with the light source $6500 K^{0}$ and ambient illumination provided only by that on the wall behind the monitor.

\subsection{Subjective Method \& Experimental Datasets}

According to ITU-R BT.2021 \cite{itu2012subjective}, there is a range of human perceptual characteristics for subjectively testing stereoscopic videos, such as video quality and visual comfort. The assessment sessions were conducted using the Double Stimulus Continuous Quality Scale (DSCQS) method with distorted and undistorted videos were played in a random order in each trial and viewers asked to record their results using different continuous scales from 0 to 5 for video quality and visual comfort. Scores close to zero indicated the worst quality or extremely uncomfortable condition whereas scores close to 5 indicated excellent video quality or a very comfortable situation.

\begin{figure}[!t]
\centering
\includegraphics[width=3.4in , height=1.2in]{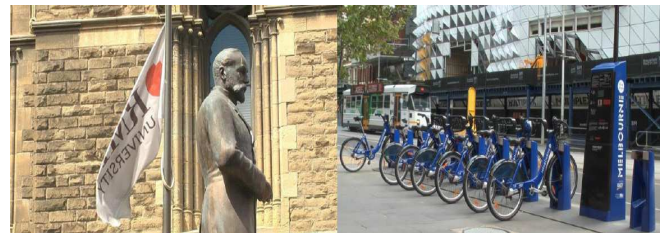}
\caption{Video datasets used for transmission and error concealment}
\label{fig_3}
\end{figure}

For our experiments, we used two different stereoscopic video sequences from RMIT3DV \cite{cheng2012rmit3dv}, each with a 10-second duration, HD resolution of $1920\times 1080$ and $25$ frames/s consisting of different types of pictorial content, such as camera and object motions, and texture as described in Table \ref{Table_1}. As shown in Fig. \ref{fig_3}, the first sequence, RMIT Statue, comprised a static close-up shot of RMIT University's statue, fluctuating flag movements and a moderate 3D effect. The second sequence, entitled Bicycle Stand consisted of a static wide shot with moving trams, different textures, and a moderate 3D effect.

\begin{table}[!t]
\renewcommand{\arraystretch}{1.3}
\caption{Experimental Video Datasets}
\label{Table_1}
\begin{center}
\begin{tabular}{lcc}  \hline \hline
 Datasets &	Disparity 	&	Video Characteristics \\ \hline \hline
 $RMIT \: Statue$	 &	$High$	 &	$Flag\: waving, very\: high\: motion $	\\
 $Bicycle \: Stand$   & $High$	 &	$Tram\: moving , variable\: motion $	\\ \hline

\end{tabular}
\end{center}
\end{table}


\subsection{Error Concealment \& Experimental Stimulus}

For simple error concealment, each pixel in a concealed frame is replaced by a corresponding pixel from a previously decoded reference frame. When a reference frame is concealed, it is both used for display and placed in the reference picture buffer for decoding subsequent pictures whereas, when a non-reference frame is concealed, the frame is used only for display. This whole procedure, which is known as the Frame Freeze approach, is still widely used in video transmission applications and, for 2D video error concealment, is termed Frame Freeze 2D. Left Frame (L.F.) Freeze defines error concealment in the left view of a 3D video while the right contains unaffected content and error concealment in both views is called Double Frame (D. F.) Freeze.

\begin{figure}[h!]
\centering
\includegraphics[width=3.5in]{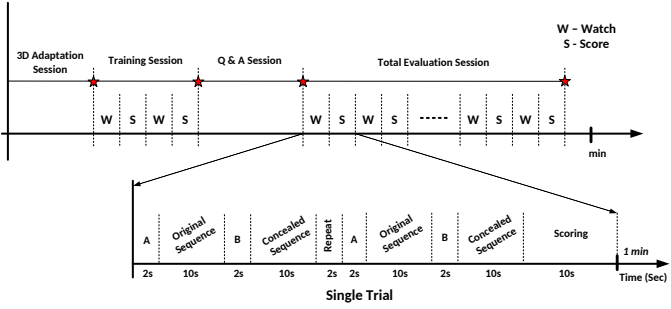}
\caption{Timelines of total evaluation session and single trial}
\label{fig_1_6}
\end{figure}

The Absolute Category Rating with the Hidden Reference (ACR-HR) method was used for the training session which was conducted before the formal evaluation so that observers would be familiar with the rating scales. Following the training session and briefing of the subjects, a question and answer session was conducted and then the video sequences evaluated and scored as shown in Fig. \ref{fig_1_6}.

The test consisted of a series of judgment trials, each involving two repetitions of two different versions of the same video material, with the first and second presentations referred to as A and B respectively, which were repeated to complete a single trial. In this study, we randomly chose either A or B as the original video sequence for comparing the performances of the different 2D and 3D video concealment methods, with two video sequences used to produce the 36 trials for our evaluation which is explained in detail at section IV.

\subsection{Analysis of Subjective Scores}

The scores given by observers are averaged to produce the Mean Opinion Score (MOS) and as shown in Equation \ref{Equ_1}:

\begin{equation}
\begin{array}{lcl}
{\hat{U}}_{jkr}=\frac{1}{N}\sum_{i=1}^{N}{U}_{ijkr}
\label{Equ_1}
\end{array}
\end{equation}

Here ${\hat{U}}_{jkr} $ is the score of observer $i$  for the degradation $j$ in video sequence $k$ and repetition $r$, and $N$ is the number of observers. To better assess the accuracy of the results, as it is desirable to combine a CI with each MOS, we used a 95\% one in

\begin{equation}
\begin{array}{lcl}
{CI}_{mean}=
\left\{\begin{matrix}
{CI}_{lower}={\hat{U}}_{jkr}-{\delta }_{jkr}
 \\
{CI}_{upper}={\hat{U}}_{jkr}+{\delta }_{jkr}
\end{matrix}\right.
\label{Equ_2}
\end{array}
\end{equation}

\begin{equation}
\begin{array}{lcl}
{\delta }_{jkr}=({T}_{val}\times {SE}_{mean})
\label{Equ_3}
\end{array}
\end{equation}

where ${\delta }_{jkr}$ is the standards deviation of mean ${\hat{U}}_{jkr}$, depend on the critical t-value (${T}_{val}$) \& standard error of the mean ${SE}_{mean}$. Standard deviation measures the amount of variability for a set of data from the mean, while the standard error of the mean measures how far the sample mean of the data is likely to be from the true population mean. For sample size $N=20$ (there were 20 viewers), the standard deviation ${S}_{jkr}$ is formulated as Equation \ref{Equ_4}.

\begin{equation}
\begin{array}{lcl}
{S}_{jkr}=\sqrt{\sum_{i=1}^{N}\frac{{({\hat{U}}_{jkr}-{U}_{jkr})}^{2}}{(N-1)}}
\label{Equ_4}
\end{array}
\end{equation}

\begin{equation}
\begin{array}{lcl}
{SE}_{mean}=\frac{{S}_{jkr}}{\sqrt{N}}
\label{Equ_5}
\end{array}
\end{equation}

As number of sample (N) is $20$, for 95\% CI and the degree of freedom of $df = N-1=19$ the critical ${T}_{val}$ is $1.729$ which is explained in details in binocular sensitivity analysis at section IV (D).

\section {Experimental Results \& Analyses}

For our experiments, we assumed that each packet was capable of carrying a complete frame and so a lost packet would lead to a full-frame loss. Since the objective of our study was to investigate the psychovisual impact on the perceived quality, the standard Frame Freeze approach was used for a 2D video. For a 3D video in which only one view was affected, the right and left views were assumed to be correctly and erroneously received respectively, with the error concealment known as the Left Frame Freeze (L. F. Freeze) and, for video in which both views were impacted; the Double Frame Freeze (D. F. Freeze) was applied. The original sequence was not revealed to the user and the order of the original and error concealed sequences varied from test to test. In this subsection, the experimental results and analysis are discussed to quantify the psychovisual impacts and enable a comparison of the 2D and 3D video error concealment methods.

Firstly, MOSs and CIs were calculated from the scores given by the subjects and then the study extended to include more analyses to investigate the psychovisual challenges of binocular sensitivity \cite{de2011sensitivity} \cite{nelson2000predicting} \cite{cogan1982monocular} and binocular rivalry between the error-concealed videos. The analyses will follow some steps to made the conclusive statements. At first, the analysis includes linear regression, to describe about 3D depth perception in brain, which enables HVS to perceive 3D videos differently than 2D. In addition, the correlation analysis will be performed between 3D and 2D to describe the sensitivity of the binocular vision and its deterioration even after same error occurrences. Finally, through the student's t-test analysis and its probability distribution we are going to determine that binocular rivalry is the main cause and significant reason of binocular vision for creating visual discomfort and lower 3D video quality even after error concealment.

\subsection{Mean Opinion Score and Confidence Interval}

For the subjective tests, 20 naive observers were asked to assess the overall quality and sensation of visual comfort associated with several pairs of video materials according to ITU-R BT.2021, with visual discomfort related to feelings of nausea, eye strain, double vision and headaches. Scoring was on continuous scales of 0 to 5 for video quality and visual comfort which ranged from worse to excellent and extremely uncomfortable to very comfortable respectively.

The widely used Frame Freeze error concealment is used for our experimental analysis. In each of the 10-second video, a single loss event is occurred and Frame Freeze error concealment is applied. The loss event can be any full frame losses of 1 to 5 frames as shown in Table \ref{Table_3} in which the L. F. Freeze and D. F. Freeze applied for error event in left view and both views respectively. Whereas random error occurrences of let view and both views are called L. F. R. Freeze and D. F. R. Freeze respectively. Each random loss was designed in such a way that there were five single-frame error occurrences with at least a 12-frame (half of the decoder content) interval between them so that viewers could properly observe every impact of the Frame Freeze approach five times. For each trial, one loss event is compared with the original version of the video. Whereas error concealment of 2D video is compared with its original 2D and error concealment of one or both views are compared with its original 3D video. As a result, each loss event consist of three trials and therefore six loss events contains total eighteen trials. For two different video datasets a subject has to attend 36 trials as shown in Table \ref{Table_3}.

\begin{table}[!h]
\caption{Comparison of 2D and 3D MOSs for different Frame Freeze concealments}
\label{Table_3}
\begin{center} {\footnotesize
\begin{tabular}{lcccc}
\hline
\hline
 CONCEALMENT & \multicolumn{2}{c}{RMIT Statue} & \multicolumn{2}{c}{Bicycle Stand} \\
Type of  & \multicolumn{1}{c}{ Video} & \multicolumn{1}{c}{Visual} &
\multicolumn{1}{c}{Video} & \multicolumn{1}{c}{Visual} \\
Frame Freeze  & \multicolumn{1}{c}{ Quality} & \multicolumn{1}{c}{Comfort} &
\multicolumn{1}{c}{Quality} & \multicolumn{1}{c}{Comfort} \\
\hline
$Original\,2D $     &     3.87 & 3.84 & 4.08 & 4.10  \\
$Original\,3D $     &     4.01	& 3.99 & 4.13 & 4.21  \\
$1\, F.\, Freeze\, 2D $        &     3.85 & 3.75 &	3.99 & 4.00 \\
$1\, L.\, F.\, Freeze $   &     3.48 & 3.34 &	3.60 & 3.60  \\
$1\, D.\, F.\, Freeze $   &     3.57	& 3.46 &	3.72 &	3.70  \\
$2\, F.\, Freeze\, 2D  $        &     3.67	& 3.67	&3.82 &	3.85  \\
$2\, L.\, F.\, Freeze $   &     3.17	& 3.12	&3.44	&3.42  \\
$2\, D.\, F.\, Freeze $   &     3.36	& 3.33	&3.51	&3.55  \\
$3\, F.\, Freeze\, 2D  $        &     3.56	& 3.59	&3.70	&3.59  \\
$3\, L.\, F.\, Freeze $   &     2.94	& 2.92	&3.20	&3.05  \\
$3\, D.\, F.\, Freeze $   &     3.23	& 3.27	&3.38	&3.27  \\
$4\, F.\, Freeze\, 2D  $        &     3.46	& 3.45	&3.45&	3.38  \\
$4\, L.\, F.\, Freeze $   &     2.80	& 2.77	&2.95	&2.75  \\
$4\, D.\, F.\, Freeze $   &     3.13 & 3.16	&3.15&	3.06  \\
$5 \, F.\, Freeze\, 2D $        &     3.29 & 3.29	&  3.30	&3.22  \\
$5\, L.\, F.\, Freeze $   &     2.59 & 2.64	&  2.78 &   2.64  \\
$5\, D.\, F.\, Freeze $   &     2.83 & 3.01   &  2.91 &   2.85  \\
$5 \, F. \, R.\, Freeze\, 2D$ &     3.55	& 3.48   &	3.55 &	3.38  \\
$5\, L.\, F.\, R.\, Freeze $  &     2.86 & 2.90    &	3.05 &	3.02  \\
$5\, D.\, F.\, R.\, Freeze $  &     3.21	& 3.19   &	3.23 &	3.10  \\
\hline
\hline
\end{tabular} }
\end{center}
\end{table}

It can be seen that, for both the RMIT Statue and Bicycle Stand video sequences, the original 3D videos obtained higher scores than the original 2D ones, which indicated that users preferred the 3D sequences. However, when a transmission loss occurred in a 3D video, it created more annoying artefacts and lower quality, as observed from the MOSs of the subjects. When the impact of frame loss was introduced, users gave the 2D videos higher scores in all cases because of the binocular sensitivity of 3D, with the HVS affected mainly by distortions caused by transmission losses. Users rated a D. F. Freeze better than a L. F. Freeze for all error events and both sequences. When only one view was frozen, binocular sensitivity was ruined and binocular rivalry occurred. This severe artefact influenced the scores for the L. F. Freeze approaches.

\begin{figure*}[!t]
\centering
\includegraphics[width=7.0in, height=4.5in]{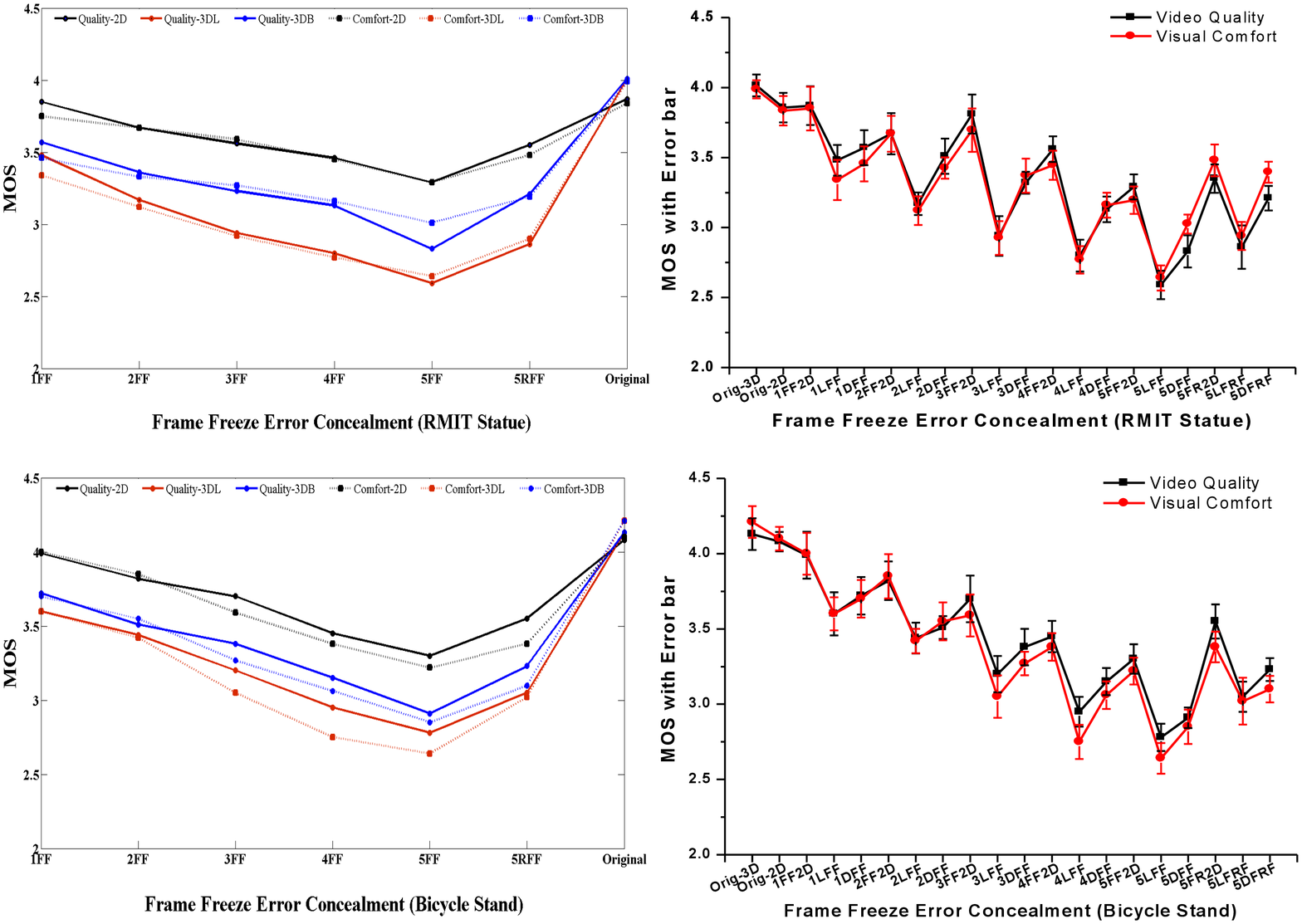}
\caption{MOS (Left) and MOS with std. error bar (right) for EC}
\label{fig_9}
\end{figure*}

Fig. \ref{fig_9} [Left] shows the MOSs for the different test scenarios for both quality (solid lines) and comfort (dashed lines), where 1R to 5R indicate concealments of 1 to 5 frames respectively and 5RR concealment of random occurrences of five single frames in a 10-second sequence. Again, it can be seen that the subjects preferred the original 3D to original 2D video. However, when transmission errors occurred and error concealment was applied, both the visual comfort and video quality scores for the 3D video decreased more quickly than those for the 2D video. The standard deviations and 95\% CIs of the MOSs is then calculated and plotted as in Fig. \ref{fig_9} [Right] in which the CIs shown as error bars with maximum value $\pm 0.15$ indicate the consistency of the MOSs acquired from the viewers.

\subsection{Binocular Vision in 3D}

A linear regression analysis \cite{yan2009linear} is performed to model the relationship between two variables by fitting a linear equation to the observed data, with one variable considered an explanatory one and the other a dependent one. It finds the straight line, called the Least Squares Regression Line (LSRL), that best represents observations in a bivariate dataset and has an equation of the form

\begin{equation}\label{eq_10}
\centering
Y = a+ bX
\end{equation}
where X is the explanatory variable, Y is the dependent variable, $a$ is a constant and $b$ is the regression coefficient. The slope of the line is $b$ and $a$ is the intercept (the value of $y$ when $x=0$).

The coefficient of determination (denoted by $R^2$) is a key output of regression analysis \cite{di2008coefficient}. It is interpreted as the proportion of the variance in the dependent variable that is predictable from the independent variable. An $R^2$ of $0.10$ means that $10$ percent of the variance in Y is predictable from X. The formula for computing the coefficient of determination for a linear regression model with one independent variable is shown in Equation \ref{eq_11}.

\begin{equation}\label{eq_11}
\centering
R^{2}=\left [ \frac{1}{N}\times \frac{\sum \left [ ( x_{i} -\bar{x})(y_{i} -\bar{y} )\right ]}{(\sigma _{x}\times \sigma _{y})} \right ]^{2}
\end{equation}
where $N$ is the number of observations used to fit the model, $\sum$ is the summation symbol, $x_{i}$ is the $x$ value for observation $i$, $\bar{x}$ is the mean $x$ value, $y_{i}$ is the y value for observation $i$, $\bar{y}$ is the mean $y$ value, $\sigma _{x}$ is the standard deviation of $x$, and $\sigma _{y}$ is the standard deviation of $y$.

\begin{figure}[!h]
\centering
\includegraphics[width=3.3 in]{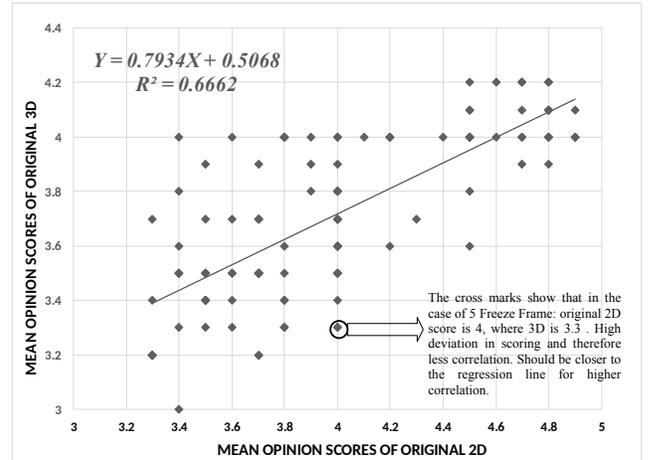}
\caption{Least Square Regression Line for monoscopic and stereoscopic videos }
\label{fig_10}
\end{figure}

We analyzed the linear regression between the MOSs obtained from both the original 2D and 3D videos that were associated with the same video dataset and the paired scores obtained for the same error concealment condition, as shown in  Fig. \ref{fig_10}. Each overlapping cross-point shows the subjective scores of the original 2D and 3D videos for a certain test case. 240 and 480 scores were obtained for the original 2D and 3D videos respectively, with 240 of the 480 3D scores used for comparison with the 120 for the monoscopic videos in order to calculate  $R^{2}$.

The calculated LSLR ($ Y=.5068+0.7934X $) shows the relationship between the monoscopic and stereoscopic videos based on the subjects' observations. Also the coefficient of determination ($R^2$) is calculated as 0.6662 using Equation \ref{eq_11} indicates the relatively low predictability of, and clear distinction between them. The decision criteria were determined based on various studies of regression which indicated that a $R^2$ value between 0.75 and 1 shows a positive to accurate correlation, 0.5 to 0.75 a weaker one and $<0.5 $ a poor association to no association ($R^2=0$) \cite{udovivcic2007we}. The $R^2$ = 0.6662 score obtained from the analysis demonstrated the non-deterministic behaviours and weaker correlations of the 2D and 3D scores, and verified that the subjects could distinguish between the same 2D and 3D videos despite the information not being disclosed to them. The regression graph confirms that, due to binocularity or the depth perception of binocular vision, the HVS perceives 2D and 3D videos differently, as indicated by variations in the MOSs of the subjects.

\begin{table*}[!h]
\centering
\caption{Binocular sensitivity - correlations of 2D \& 3D error concealments for Bicycle Stand video sequence}
\label{tab_4}
\renewcommand{\arraystretch}{1.5}%
\begin{tabular}{|l|ccccccc|}
\hline
\textbf{Correl} & \textbf{Org\_2D} & \textbf{Org\_2D} & \textbf{1F\_2D} & \textbf{2F\_2D} & \textbf{3F\_2D} & \textbf{4F\_2D} & \textbf{5F\_2D} \\ \hline
\textbf{Org\_2D} & 1 & - & - & - & - & - & - \\
\textbf{Org\_3D} & - & 0.820 & - & - & - & - & - \\
\textbf{1LF\_3D} & - & - & 0.801 & - & - & - & - \\
\textbf{2LF\_3D} & - & - & - & 0.772 & - & - & - \\
\textbf{3LF\_3D} & - & - & - & - & 0.704 & - & - \\
\textbf{4LF\_3D} & - & - & - & - & - & 0.635 & - \\
\textbf{5LF\_3D} & - & - & - & - & - & - & 0.598 \\ \hline
\end{tabular}
\end{table*}

\begin{table*}[]
\centering
\caption{Binocular rivalry - correlations of error concealments in single and double views for Bicycle Stand video sequence}
\label{tab_5}
\renewcommand{\arraystretch}{1.5}%
\begin{tabular}{|c|lccccc|}
\hline
\textbf{Correl.} & \multicolumn{6}{c|}{\textbf{Error Concealment (EC) in Double-view 3D}} \\ \hline
 & \textbf{Methods} & \multicolumn{1}{l}{\textit{1DF\_3D}} & \multicolumn{1}{l}{\textit{2DF\_3D}} & \multicolumn{1}{l}{\textit{3DF\_3D}} & \multicolumn{1}{l}{\textit{4DF\_3D}} & \multicolumn{1}{l|}{\textit{5DF\_3D}} \\
\textbf{EC} & \textit{1LF\_3D} & 0.805 & - & - & - & - \\
\textbf{in} & \textit{2LF\_3D} & - & 0.764 & - & - & - \\
\textbf{Single-view} & \textit{3LF\_3D} & - & - & 0.722 & - & - \\
\textbf{3D} & \textit{4LF\_3D} & - & - & - & 0.656 & - \\
\multicolumn{1}{|l|}{} & \textit{5LF\_3D} & - & - & - & - & 0.615 \\ \hline
\end{tabular}
\end{table*}

\subsection{Impact of Binocular Sensitivity and Rivalry }

The next analysis was performed to calculate the correlations between the 2D and 3D error concealment approaches to determine the effect of binocular sensitivity. Then, those between the single-view and double-view error concealment strategies were analyzed to measure the impact of binocular rivalry. A mathematical relationship between two datasets can be represented using the correlation coefficient, as shown in Equation \ref{eq_12}, which gives an output range from +1 to -1 to indicate a positive correlation (close to +1) through no correlation (close to zero) to a negative one (close to -1) \cite{liang2008analysis}.

\begin{equation}\label{eq_12}
\centering
Correl(X,Y) = \frac{\sum(x-\overline{x})(y-\overline{y})}{\sqrt{\sum (x-\overline{x})^{2} \sum (y-\overline{y})^{2}}}
\end{equation}

In Table \ref{tab_4}, we can see that the binocular impact increases when distortions raise from one to five frames after error concealment, as reflected in the MOSs. Also, using Equation 12 to determine the correlation coefficient values, it is clear that the correlation between the 2D and 3D videos decreased with increasing numbers of error events. Initially, the correlation between the original 2D (Org\_2D) and original 3D (Org\_3D) was high and, after error concealment for frame losses, began to decrease rapidly. It can be observed that the correlation between 1F\_2D and 1F\_3D was $0.801$ and whereas between 2F\_2D and 2F\_3D it is $0.772$. However, it might be expected, it should have been the same in all cases as the same error frames and therefore the same distortions happened in both 2D and 3D videos for any particular error event. This strongly indicated that 3D binocular vision was very sensitive and significantly affected as the number of frozen frames increased.

A second correlation analysis of single- and double-view error concealments was performed to estimate only the effect of binocular rivalry without that of binocular sensitivity. The results in Table \ref{tab_5} show that the correlations between the methods decreased and those between their binocular rivalry artefacts increased as the lengths of the numbers of error events increased. From the comparisons in Tables \ref{tab_4} and \ref{tab_5}, it is clear that, as the number of error events in the 2D and 3D videos increased, the correlations between them decreased. If the deterioration is increased from 1F\_2D to 2F\_2D, the same drop also happened in 2F\_3D from 1F\_3D. Although, ideally, the correlations between 2D and 3D should be the same in a particular event, the experimental results showed that they decreased in proportion to the number of error events. The reasons for these outcomes were the disruption of binocular sensitivity and increase in the impact of binocular rivalry between the error concealments in the single (L. F.) and both (D. F.) views.

Through the above sensitivity analysis it is observed that 3D binocular vision (depth perception) is affecting the correlation between 2D and 3D video. In addition, lower opinion scores in Table \ref{Table_3} shows that single view 3D error concealment is the main cause of this effect. In the next section we further analyze the subject results using a student t-test to determine the statistical significance of our results.

\subsection{Binocular Sensitivity in 3D Error Concealment}

\begin{figure}[!h]
\centering
\includegraphics[width=3.4in]{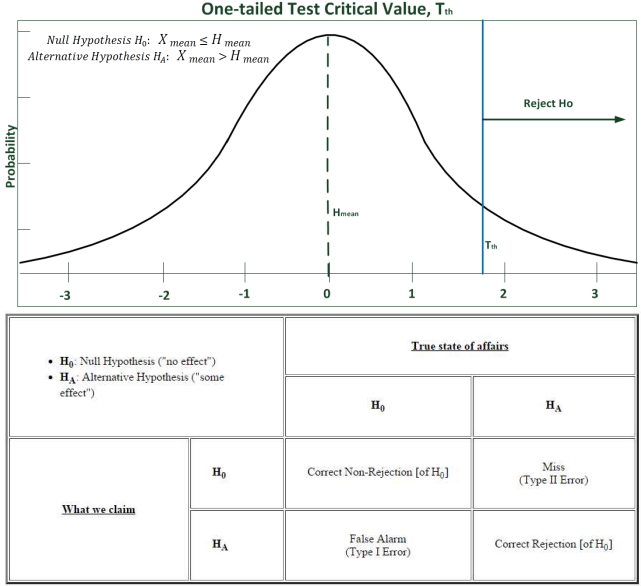}
\caption{Distribution of one-tailed dependent samples (paired t-test) and corresponding hypothesis \cite{statistics} (erroneous rejection of ${H}_{0}$  (although true) constitutes Type 1 error and erroneous acceptance of ${H}_{0}$  (although false) Type 2 error}
\label{fig_8}
\end{figure}

In our approach, we used the paired-samples t-test \cite{sprinthall2011basic} \cite{ttest} to compare the means of two variables for a single group by computing the differences between the values of these variables for each case and testing whether the average difference was significantly different from zero. This sample was then used to compare the mean of a single sample of scores with a known or hypothetical population mean $({H}_{mean})$ as shown in Fig. \ref{fig_8}.

Based on the null hypothesis, we computed whether one group of samples was better than the other through a t-probability distribution, where $N$ was the equal number of samples in each distribution and ${S}_{1},{S}_{2} $ were the two sample groups. The student t-test, ${T}_{test}$  was  used to express the CIs for a set of data and compare the results obtained in different experiments. The aim was to state the possible range of true mean,${\hat{U}}_{true}$, from the measured mean ${\hat{U}}_{jkr}$ with probability.

\begin{equation}
\begin{array}{lcl}
{\hat{U}}_{true}={\hat{U}}_{jkr}\pm {T}_{test}\frac{{S}_{jkr}}{\sqrt{N}}
\label{Equ_6}
\end{array}
\end{equation}

\begin{equation}
\begin{array}{lcl}
T_{test}=\frac{X_{mean} - H_{mean}}{\sqrt{\frac{\sum_{i=1}^N (X_{i} - X_{mean})^{2}}{N(N-1)}}}
\label{Equ_7}
\end{array}
\end{equation}

where $X_i$ represents the differences between sample groups $S_1$  and $S_2$, and $X_{mean}$ the mean of these differences $(X_{mean}=\frac{X_{i}}{N}=\frac{\sum_{i=1}^N(S_{1i}-S_{2i})}{N})$. The calculated student's t-test value ${T}_{test}$  was compared with the critical (threshold) value ${T}_{th}$  corresponding to the degree of freedom $(df)$ calculated as $df= N-1 $ and then the confidence level, i.e., 90\% or 95\% was chosen. If ${T}_{test}\geq {T}_{th}$, ${H}_{0}$ was rejected, otherwise ${H}_{0}$ was accepted. The critical threshold ${T}_{th}$ was estimated from the probability of obtaining test statistics called P-value. The obtained ${T}_{test}$ value used to calculate the P-value to determine how confident we are that the difference between the two distributions was significant. Statistically, the P-value is the smallest observed significance level for which the null hypothesis is rejected. That is, the smaller it was, the more evidence there was to support the alternative hypothesis.

In the interval $(a, b)$, the area under the t-probability distribution required the value of the following integral where the degree of freedom to be calculated, $df=19$ and $\Gamma$ the gamma function defined by the integral \cite{sprinthall2011basic} was

\begin{equation}
\begin{array}{lcl}
P-Value = \int_{a}^{b}f({T}_{test})d{T}_{test}
\label{Equ_8}
\end{array}
\end{equation}

where $f({T}_{test})$ is defined by:\newline

\begin{equation}
\begin{array}{lcl}
f({T}_{test}) = \frac{\Gamma ( \frac{df+1}{2})}{\sqrt{df\Pi} \Gamma (\frac{df}{2})}{(1+\frac{{{T}_{test}}^{2}}{df})}^{-(\frac{df+1}{2})}
\label{Equ_9}
\end{array}
\end{equation}

Hypothesis tests are used to test the validity of a claim made about a population. In this experiment, the proposed method $S_1$  was compared with existing methods as $S_2$. We analyzed the t-test scores from Equation \ref{Equ_8} and calculated the P-value using Equation \ref{Equ_9}. Assuming that the proposed approach was worse than existing concealment ones, the new population mean $(X_{mean})$ should have been less than or equal to the hypothetical population mean $(H_{mean})$ as follows:
\newline

\begin{quote}
  \centering
 \fontsize{10.5}{10}\selectfont{\textbf{ Null Hypothesis $H_0:X_{mean} \leq H_{mean}$}} \\
 \fontsize{10.5}{10}\selectfont{\textbf{Alternative Hypothesis $H_A:X_{mean} > H_{mean}$}}\newline

 \end{quote}

Results supporting the alternative hypothesis indicated that the proposed method was better than existing approaches. Because we were interested in only one side of the distribution, a one-sided or one-tailed test was applied to calculate the t-test value. As we took the difference between two populations, the mean of this difference should have equalled $0$ as our hypothetical mean. For a 95\% CI, the P-value should have been $\leq 0.05$ and, for $df = 19$, $T_{test}$ needed to be $\leq 1.729$ to accept the null hypothesis, otherwise, the alternative hypothesis would be accepted. As a small P-value (typically $\leq 0.05$ ) provided strong evidence against the null hypothesis, as this hypothesis could be rejected, there would be a statistically significant difference between the two populations which would help the understanding of and prove the difference between these populations. Based on different P-values, statistical significance was classified in the several categories shown in Table \ref{tab_2}.

\begin{table}[!h]
\centering
\caption{Categories of statistical significance based on P-value}
\label{tab_2}
\renewcommand{\arraystretch}{1.5}%
\begin{tabular}{cccc}
\hline
\multicolumn{1}{c}{\textbf{Type}} & \textbf{Significance} & \textbf{P-value} & \textbf{Comment} \\
\hline
\textbf{ExSS} & \begin{tabular}[c]{@{}c@{}}Extreme Statistical\\ Significance\end{tabular} & \textbf{$\leq 0.001$} & \begin{tabular}[c]{@{}c@{}}Extremely strong evidence \\ against the null hypothesis\end{tabular} \\ \hline
\textbf{VSS} & \begin{tabular}[c]{@{}c@{}}Very Statistically\\ Significant\end{tabular} & \textbf{$\leq 0.01$} & \begin{tabular}[c]{@{}c@{}}Very strong evidence \\ against the null hypothesis\end{tabular} \\ \hline
\textbf{SS} & \begin{tabular}[c]{@{}c@{}}Statistically\\ Significant\end{tabular} & \textbf{$\leq 0.05$} & \begin{tabular}[c]{@{}c@{}}Strong evidence \\ against the null hypothesis\end{tabular} \\ \hline
\textbf{NqSS} & \begin{tabular}[c]{@{}c@{}}Not Quite \\Statistically \\Significant\end{tabular} & \textbf{$\leq 0.1$} & Marginal \\ \hline
\textbf{NSS} & \begin{tabular}[c]{@{}c@{}}Not Statistically \\Significant \end{tabular} & \textbf{$> 0.1$} & Fail to reject null hypothesis \\ \hline
\end{tabular}
\end{table}

\begin{table*}[t]
\caption{Statistical significance of video quality and visual comfort for Frame Freeze error concealments}
\label{Table_6}
\begin{center}
\renewcommand{\arraystretch}{1.5}%
{\footnotesize
\begin{tabular}{l|l|crccrc|crccrc}

\hline
\multicolumn{2}{c|}{\small{\textbf{Perception $\Rightarrow$}}} & \multicolumn{6}{c|}{\small{\textbf{Video Quality}}} & \multicolumn{6}{c}{\small{\textbf{Visual Comfort}}} \\\hline
\multicolumn{1}{l|}{Video} & \multicolumn{1}{c|}{Compare} & \multicolumn{3}{c}{Left Frame Freeze\:(3D)} & \multicolumn{3}{c|}{Double Frame Freeze\:(3D)} & \multicolumn{3}{c}{Left Frame Freeze\:(3D)} & \multicolumn{3}{c}{Double Frame Freeze\:(3D)} \\
\multicolumn{1}{c|}{Dataset} & \multicolumn{1}{c|}{ with 2D} & \multicolumn{1}{c}{$T_{test}$} & \multicolumn{1}{r}{P-val} &\multicolumn{1}{c}{S. Sig.} & \multicolumn{1}{c}{$T_{test}$} & \multicolumn{1}{r}{P-val} &\multicolumn{1}{c|}{S. Sig.} & \multicolumn{1}{c}{$T_{test}$} & \multicolumn{1}{r}{P-val} &\multicolumn{1}{c}{S. Sig.} & \multicolumn{1}{c}{$T_{test}$} & \multicolumn{1}{r}{P-val} &\multicolumn{1}{c}{S. Sig.}\\ \hline

&1 Frame &4.319&.00020   &ExSS& 2.027&.02850   &SS  &5.444&.00001   &ExSS&4.532&.00001   &ExSS\\
RMIT
&2 Frames&4.825&.00007   &ExSS& 2.954&.00410   &VSS &7.193&$<.00001$&ExSS&5.059&.00003   &ExSS\\
Statue
&3 Frames&9.649&$<.00001$&ExSS& 6.090&$<.00001$&ExSS&6.447&$<.00001$&ExSS&5.365&.00002   &ExSS\\
&4 Frames&15.49&$<.00001$&ExSS& 12.43&$<.00001$&ExSS&9.266&$<.00001$&ExSS&5.667&$<.00001$&ExSS \\
&5 Frames&21.21&$<.00001$&ExSS& 15.79&$<.00001$&ExSS&16.17&$<.00001$&ExSS&7.072&$<.00001$&ExSS \\\hline
&1 Frame &3.284&.00246   &VSS	& 2.681&.00946 &VSS &4.013&.00104   &VSS &2.862&.00660   &VSS\\
Bicycle
&2 Frames&3.597&.00102   &VSS	& 2.755&.00365 &VSS &4.893&.00007	&ExSS&3.335&.00205   &VSS\\
Stand
&3 Frames&7.129&$<.00001$&ExSS& 7.090&$<.00001$&ExSS&6.887&$<.00001$&ExSS&4.985&.00005   &ExSS\\
&4 Frames&10.23&$<.00001$&ExSS& 9.330&$<.00001$&ExSS&9.422&$<.00001$&ExSS&7.121&$<.00001$&ExSS\\
&5 Frames&15.17&$<.00001$&ExSS& 11.11&$<.00001$&ExSS&12.72&$<.00001$&ExSS&10.01&$<.00001$&ExSS\\
\hline
\end{tabular} }
\end{center}
\end{table*}

As the HVS is more sensitive to binocular than monocular vision, it is also more sensitive to error concealment in 3D than 2D video. If there is any mismatch between the contrast, color, disparity, shape, etc., perceived by each eye, this creates both significant visual discomfort and a lower perceived video quality. Through the MOSs, it was demonstrated that, after error concealment, viewers rated the quality and comfort of a 3D video lower than those of a 2D one with the same level of impairment (Figure \ref{fig_8}), results we then analyzed statistically. We calculated the student's t-test scores  ($ {T}_{test}$ ) scores to determine the statistical significance of binocular sensitivity to 3D by comparing 2D Frame Freeze approaches with 3D L. F. Freeze and D. F. Freeze ones. The P-value scores from Equations 8 and 9 were analyzed to determine the statistical significance of the 2D error concealment methods compared with those of the single-view (L. F. Freeze) and double-view (Double Freeze) 3D error concealment methods, as shown in Table \ref{Table_6} in terms of video quality and visual comfort respectively. As explained earlier, a one-tailed paired (dependent) t-test was performed using 20 subjects to calculate the statistical significance considering the null hypothesis, which is: \newline

$H_0$: 2D error concealment is worse than stereoscopic error concealment in one or both views. \newline

For a 95\% CI, the P-value should be at least $\leq 0.05$ and the degree of freedom (df) $(N-1)=19$, with the T\_test needing to be $\leq 1.729$ to accept the null hypothesis. Otherwise, $H_0$ must be rejected and the alternative hypothesis accepted. The statistical significance categories were ranked as in Table \ref{tab_2}. Based on the null hypothesis and from Table \ref{Table_6}, we can observe that in all cases the null hypothesis is strongly rejected which implied that error concealment in a 2D video was far better than that in a 3D one in terms of both visual comfort and video quality. These statistical significance analyses also demonstrated that the deteriorating correlation (Table \ref{tab_4}) was due mainly to the binocular vision and sensitivity for a 3D video. Note that, of 20 comparisons of observers, extremely statistically significant results were achieved for 17 for L. F. Freeze and 14 for D. F. Freeze. It was clear that binocular sensitivity had a significant impact after 3D error concealment. The statistics in \ref{Table_6} also indicate that the L. F. Freeze impact, referred to as that of binocular rivalry, was very severe for not only the visual comfort level but also the deteriorating 3D video quality as the null hypothesis was heavily rejected because it obtained either higher t-test scores or lower P-values (worse condition) than D. F. Freeze in all cases. The different impacts of binocular rivalry in 2D and 3D videos are examined in detail in the next subsection.

\begin{table*}[!t]
\centering
\caption{Statistical significance of single-view and double-view Frame Freeze in 3D error concealments}
\label{tab_7}
\renewcommand{\arraystretch}{1.5}%
\begin{tabular}{|l|c|c|c|c|c|c|c|}
\hline
\multicolumn{2}{|c|}{\small{\textbf{Perceptual Dimensions $\Rightarrow$}}} & \multicolumn{3}{c|}{\small{\textbf{Video Quality}}} & \multicolumn{3}{c|}{\small{\textbf{Visual Comfort}}} \\ \hline
\textbf{\begin{tabular}[c]{@{}l@{}}Video Datasets\end{tabular}} & \textbf{\begin{tabular}[c]{@{}l@{}}Frame Freeze \\Error Concealment\end{tabular}} & $T_{test}$ & \textbf{P-val} & \textbf{Stat.Sig} & $T_{test}$ & \textbf{P-val} & \textbf{Stat.Sig} \\ \hline
 & \textbf{1 Frame}  & 1.4954 & 0.07570   & NqSS & 2.1924 & 0.02050 & SS \\ \cline{2-8}
 & \textbf{2 Frames} & 3.8231 & 0.00057   & ExSS & 4.5445 & 0.00011 & ExSS \\ \cline{2-8}
 {\small{\textbf{RMIT Statue}}}
 & \textbf{3 Frames} & 5.8751 & $<.00001$ & ExSS & 4.9787 & 0.00004 & ExSS \\ \cline{2-8}
 & \textbf{4 Frames} & 5.8762 & $<.00001$ & ExSS & 7.6875 & $<.00001$ & ExSS \\ \cline{2-8}
 & \textbf{5 Frames} & 3.7848 & 0.00063   & ExSS & 8.1562 & $<.00001$ & ExSS \\ \hline
 & \textbf{1 Frame}  & 1.7633 & 0.18340   & NSS  & 2.0046 & 0.02990 & SS \\ \cline{2-8}
 & \textbf{2 Frames} & 2.0544 & 0.03860   & SS   & 3.5685 & 0.00497 & VSS \\ \cline{2-8}
 {\small{\textbf{Bicycle Stand}}}
 & \textbf{3 Frames} & 4.5860 & 0.00008   & ExSS & 5.5942 & $<.00001$ & ExSS \\ \cline{2-8}
 & \textbf{4 Frames} & 4.8661 & 0.00006   & ExSS & 6.8753 & $<.00001$ & ExSS \\ \cline{2-8}
 & \textbf{5 Frames} & 5.7514 & $<.00001$ & ExSS & 6.9482 & $<.00001$ & ExSS \\ \hline
\end{tabular}
\end{table*}

\subsection{Effect of Binocular Rivalry in 3D Error Concealment}

It might be expected that the impact of error concealment in both views would be worse than for error concealment of a single view. However, the lower MOSs for L. F. Freeze than D. F. Freeze revealed the influence of another aspect of binocular vision which, according to vision research \cite{zhang2011binocular}, \cite{lei1994spatial}, \cite{wang2011considering}, \cite{howard1995binocular} is called binocular rivalry - a visual phenomenon that occurs when dissimilar monocular stimuli are presented to both eyes, as discussed earlier. The best way of proving this effect was to perform a statistical comparison of error concealments in one and two views of a stereoscopic video, a Student t-test procedure was conducted followed by the same one-tailed paired t-test approach with the null hypothesis that:\newline

$H_0$: For stereoscopic video, error concealment in both views is worse than that in one view.\newline

The results from the statistical significance analysis of error concealments in one and both views are shown in \ref{tab_7} with their calculated t-test and P-value scores and different significance conditions allocated based on the threshold levels defined in \ref{tab_2}. From the statistical significance levels, we can observe the psychovisual differences in the MOSs caused by the binocular rivalry in 3D depth perception. It is clear that, for the RMIT Statue and Bicycle Stand datasets, at least statistically significant (SS) differences were obtained in 18 of the 20 cases (all in visual comfort), of which 14 were extremely statistically significant (ExSS). These SS or ExSS implied that the null hypothesis can be rejected indicating that error concealments in both views were far better than those in one view. By definition, this meant that binocular rivalry was a significant psychovisual impact affecting error concealment in a single view. Also, for video quality, statistical significance was achieved for 8 of 10 cases in this test and, even in the other two, a D. F. Freeze was still better than a single one, as indicated by the MOSs in Table \ref{Table_3}, an effect which became stronger as the number of frame freezes increased.

\section{Conclusion}
3D technology and its display will depend crucially on how realistically natural viewing can be achieved after error concealment. A great deal of research has led to a good understanding of the performance of the HVS when presented with 2D video material. The purpose of this study was to identify the binocular sensitivity and binocular rivalry effects of applying error concealment and demonstrate that the simple approaches used for 2D videos are not suitable for 3D ones without appropriate modifications. All of our experiments verified that binocular rivalry had an enormous impact on the perceived video quality and visual comfort of a viewer while watching stereoscopic videos and the stepwise breakdown of experimental analyses led to the following conclusive statements that need to be considered carefully before designing 3D error concealment approaches.

\begin{enumerate}
  \item Binocular vision in 3D: the HVS can clearly perceive 2D and 3D videos separately due to depth perception.
  \item Impact of binocular vision: there is a correlation that exists between 2D and 3D approaches. However, this correlation decreases when error concealment is introduced which affects the sensitive binocular vision.
  \item Binocular sensitivity: 3D error concealment is more sensitive to the HVS than the same error concealment in 2D due to its greater binocular impact.
  \item Binocular rivalry is a significant psychovisual challenge for asynchronous and distorted views in stereoscopic video. Experimental results and comparative analyses between single-view and double-view 3D error concealments techniques showed binocular rivalry is the cause of visual discomfort for 3D binocular vision.
\end{enumerate}

Several statistical analyses of the severity of binocular impacts (sensitivity and rivalry) in 3D videos using an existing error concealment technique for video transmission were performed. However, using this simple method, it was found that, due to binocular rivalry, 3D videos were more distorted than 2D ones from a viewers' perspective and this could affect the HVS and cause eye strain, headaches, nausea, fatigue, etc. These findings will help us to design a new stereoscopic video transmission and decoding system and establish appropriate standard subjective evaluation criteria for measuring the quality and visual comfort of 3D videos. An important step towards achieving this is to develop a real-time error concealment strategy widely applicable for all types of 3D videos, which does not negatively affect a user's view and ensure visual comfort. This is a topic for our future work.

\balance
\bibliographystyle{IEEEtran}
\bibliography{ref}
\end{document}